\documentclass[twocolumn, prd, aps, nofootinbib,superscriptaddress]{revtex4-1}

\usepackage{graphicx}  
\usepackage{amssymb}   
\usepackage{amsfonts}       
\usepackage{mathrsfs} 
\usepackage{amsmath}
\usepackage{amssymb}
\usepackage{graphicx}
\usepackage{hyperref}
\usepackage{tikz} 
\usepackage{color}
\usepackage{empheq}
\begin{document}



\title{Spicing up the recipe for echoes from exotic compact objects: orbital differences and corrections in rotating backgrounds}
\author{Lu\'is Felipe Longo Micchi}
\email[]{luis.longo@ufabc.edu.br}
\affiliation{Centro de Ci\^encias Naturais e Humanas, UFABC, Santo Andr\'e, SP  09210-170, Brazil}
\author{Cecilia Chirenti}
\email[]{cecilia.chirenti@ufabc.edu.br}
\affiliation{Centro de de Matem\'atica, Computa\c c\~ao e Cogni\c c\~ao, UFABC, Santo Andr\'e, SP  09210-170, Brazil}
\affiliation{Department of Astronomy, University of Maryland, College Park, MD 20742-2421, USA}
\affiliation{Center for Research and Exploration in Space Science and Technology, NASA/GSFC, Greenbelt, MD 20771, USA}

\begin{abstract}

Recently it has been argued that near-horizon modifications of the standard (classical) black hole spacetime could lead to observable alterations of the gravitational waveform generated by a binary black hole coalescence. Such modifications can be inspired by quantum gravity considerations, resulting in speculative horizonless exotic compact objects (ECOs) with no singularities, which may be an alternative to the classical black hole solution. A largely model-independent description of these objects proposed in the literature relies on the introduction of a partially reflective wall at some small distance away from the ``would-be" horizon. 
The inspiral-merger-ringdown of a pair of such objects would be subject to possibly detectable deviations from the black hole case due to matter effects. In particular, the ringdown phase would be modified by the late emergence of so-called ``echoes" in the waveform, but most studies so far have considered spherically symmetric backgrounds. We use an in-falling scalar charge as an initial perturbation to simulate the excitation of the echoes of a rotating ECO and we explore both the co-rotating and counter-rotating cases, which provide distinct signals.
In particular, rotation breaks the symmetry between positive and negative frequencies and introduces a subdominant frequency contribution in each echo, which we examine here for the first time. Our results follow consistently from the solution of the Teukolsky equation using the MST method developed by Mano, Suzuki and Takasugi, and the construction of the complex Green's function integrated over different particle geodesics.
\end{abstract}

\pacs{\textcolor{red}{add PACS}}
\maketitle


\section{Introduction}
\quad The existence of black holes is supported by a large number of astrophysical observations of compact objects with about ten solar masses in binary systems, mostly in our galaxy, as well as supermassive compact objects in the center of galaxies, ranging typically from $10^6-10^9 M_{\odot}$. More recently, the strongest evidence for the existence of black holes comes from the detections of gravitational waves by the LIGO-Virgo collaboration \cite{Ligo1,Ligo2,Ligo3,Ligo4,Ligo5,Ligo6,Ligo7}, supplemented by the results of the Event Horizon Telescope collaboration \cite{EHT1,EHT2,EHT3,EHT4}. 

However, there is still room for alternative theories and models despite the increasingly more stringent tests and repeated successes of the predictions of general relativity. Moreover, from the point of view of a future theory of quantum gravity it is quite possible that the classical black hole picture \emph{must} be replaced by a different solution. The most particular and distinguishable feature of a black hole in general relativity  is the presence of an event horizon, but finding irrefutable proof of its existence might be a nearly impossible task \cite{Abramowicz}. 

Different alternative models have been suggested in the literature as ``black hole mimickers'', that is, horizonless compact objects that could in principle reproduce the observed black hole phenomenology without having a horizon or a singularity. Examples include gravastars \cite{Gravastar,Mazur9545}, boson stars \cite{BosonStar}, 2-2 holes \cite{22hole}, firewalls and fuzzballs \cite{Fuzzball}.

The characteristics and possible observational signatures of such ``black hole alternatives", also called exotic compact objects (ECOs), have been extensively studied (see \cite{cardoso2019testing} for a review). In compact binary coalescences, which are usually described in terms of inspiral, merger and ringdown, the inspiral gravitational waveform is insensitive to the character of the compact objects to leading order\footnote{But finite size effects come in at the 5PN, which yielded interesting constraints on the neutron star equation of state from the event GW170817 \cite{LigoEOS}. ECOs will also have an equation of state!}, although it has been recently argued that constraints from tidal heating could be obtained in extreme mass ratio inspirals \cite{Tidal}. 

The ECO merger waveform would have to be determined through numerical relativity simulations, which would need either a concrete realization of an ECO model or substantial adaptations of the boundary conditions close to but still outside the black holes' horizons. 

During the ringdown after the merger, the linear perturbation regime is recovered and the waveform can be described in terms of characteristic exponentially damped oscillations, the quasinormal modes (QNMs) of the spacetime \cite{Kokkotas1999,Berti_2009}. Therefore the ECO spacetime will have different QNMs from a black hole spacetime (see for example \cite{2006CQGra..23.2303D,Chirenti2,2009PhRvD..80l4047P}), and the detection of unexpected frequencies in a gravitational wave signal would be a clear indication of a deviation from the black hole metric. In \cite{Chirenti1} it was shown that the quasinormal modes of a realization of the gravastar model fail to match the observed fundamental ringdown frequency and damping time in GW150914.

\begin{figure}
	   \includegraphics[width=0.5\textwidth]{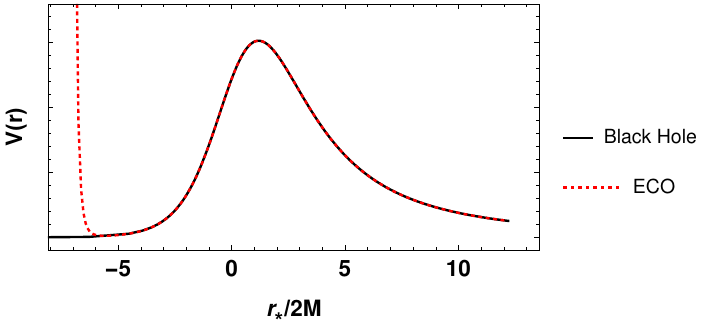}
\caption{Schematic representation of the effective scattering potential for a black hole and for a very compact star, or an ECO. In the ECO case, an incoming wave can be partially trapped in the potential well and repeated reflections will generate the echoes in the waveform seen by a far away observer. The tortoise coordinate $r_*$ is defined in eq. (\ref{eq:r*}).}
\label{Fig1}
\end{figure}

But the structure of the ringdown signal can become more complex for increasingly more compact objects, with a series of subsequent exponentially damped oscillations which were first calculated for uniform density stars in \cite{Ferrari} and later called ``echoes''  by \cite{Cardosoprobehorizon} in the context of ECOs. This behavior comes from the repeated scattering of the waves in the effective potential well formed by the peak of the potential at the photon ring and the centrifugal barrier near the center of the star, as shown schematically in Figure \ref{Fig1}.

This led to the proposal in \cite{Cardosoprobehorizon} that the \emph{early} ringdown of an ECO merger would result from the scattering on the peak and be indistinguishable from the black hole quasinormal mode signature if the surface of the ECO is infinitesimally close to the event horizon of a black hole with the same mass. In this scenario, it would be possible to distinguish an ECO from a black hole only during the \emph{later} part of the ringdown, when there would have been enough time for the characteristic frequencies of the ECO to emerge in the echoes. The existence of such pulses in gravitational wave detections would imply an enormous shift in our current understanding of black holes and general relativity. 

Consequently, a lot of recent effort has been put into the modeling of these signals using different approaches. Initially only nonrotating configurations were considered, as in \cite{Cardosoprobehorizon}. In \cite{Mark:2017dnq} scalar echoes were studied on a non-rotating ECO background using an infalling particle as an initial excitation. Gravitational echoes from the head-on collision of two boson stars were examined in \cite{EchoesofECOS}.

Adding rotation to horizonless compact objects is a nontrivial issue due to the ergoregion instability \cite{1978RSPSA.364..211C}. In \cite{2008PhRvD..77l4044C,Chirenti3} the ergoregion instability of rotating gravastars was considered for scalar perturbations, showing the limitations in the allowed parameter space of the solutions to prevent the occurrence of the instability. Later, \cite{Paniquench,RoleofAbsorption} showed that the instability can be effectively suppressed if the ECO has a small absorption rate. Scalar echoes from rotating wormholes were calculated in \cite{Buenokerrlike}.

Different works presented proposals for echo templates that could be used in the search for echoes in gravitational wave data (see for example \cite{PaniAnalytical,Cardosomorphology,AfshordiManual}). Some preliminary searches reported evidence for echoes with low significance $< 2.5\sigma$ \cite{AfshordiDetection,Holdomnewwindows}, which were not confirmed by other studies \cite{WesterweckDetection,Ricotemplate,tsang2019morphologyindependent}.

Here we use an infalling scalar point charge to excite the scalar echoes of a rotating ECO. We follow the semi-analytical method developed by Mano, Suzuki, and Takasugi (the MST method) \cite{MSToriginal, SasakiReview} to solve the Teukolsky equation and obtain the time-dependent waveform. We find that rotation breaks the symmetry between positive and negative frequency contributions, making the echo signature dependent on the orbital motion followed by the infalling particle, and also introducing a subdominant frequency contribution in each echo.

The structure of this paper is as follows: in Section \ref{constructing} we present the mathematical formulation used in our work and we show some intermediate results necessary for the construction of the echoes. In Section \ref{sect6} we present the echo waveforms for rotating ECOs and discuss how they can be affected by different geodesic motions of the infalling particle. Finally, in Section \ref{sect7} we state our conclusions. We use $c = G = 1$ units throughout this work.

\section{Constructing echoes from rotating ECO's}
\label{constructing}

\subsection{ECO Model}
\label{sec:model}

We are interested here in objects that are almost as compact as black holes, such that the exterior spacetime geometry can be well approximated by the Kerr metric, given in Boyer-Lindquist coordinates as 
\begin{align}
&ds^{2} = -\dfrac{\Delta}{\Sigma} (dt-a \sin(\theta)d\phi)^{2} \nonumber +\\&+\dfrac{\sin^{2}(\theta)}{\Sigma} \left((r^{2}+a^{2})d\phi-adt\right)^{2}  + \dfrac{\Sigma}{\Delta}dr^{2} + \Sigma d\theta^{2} ,
\label{eq:Kerr}
\end{align}
where $\Delta\equiv r^{2} -2Mr +a^{2}$, $\Sigma=r^{2}+a^{2}\cos^{2}(\theta)$, $a$ and $M$ are the spin and mass of the black hole, respectively. This metric possesses two horizons, an event horizon at $r=r_{+}\equiv M+\sqrt{M^{2}-a^{2}}$ and a Cauchy horizon at $r=r_{-}\equiv M-\sqrt{M^{2}-a^{2}}$. 

Although there is no equivalent of Birkhoff’s theorem for axially symmetric spacetimes, some recent works have shown that the normalized moment of inertia and the mass quadrupole moment of a slowly rotating compact star approach the corresponding values for the Kerr metric, see for example \cite{PosadaMomentumInertia}. Therefore we will describe the spacetime exterior to the ECO by the standard Kerr metric (\ref{eq:Kerr}).

In order to model this horizonless spacetime, we will follow a prescription initially suggested in \cite{Mark:2017dnq} and suppose the existence of a reflective wall, which will encode the characteristics of the surface and interior of the ECO. This wall is placed on the surface of the object at a radial coordinate $r_0$ just above the expected position for the event horizon $r_+$, that is, $r_{0}= r_{+}+\delta$, where $\delta\ll M$. It is useful to define the position of the wall in terms of the tortoise coordinate $r_*$, given as usual by 
 \begin{equation}
    \dfrac{d r_{*}}{dr} = \dfrac{r^{2}+a^{2}}{\Delta}\,,
    \label{eq:r*}
\end{equation}
and we choose the integration constant to have 
\begin{displaymath}
    r_{*}= r + \dfrac{ 2Mr_{+}}{r_{+}-r_{-}}\ln \left(\frac{r-r_{+}}{2M}\right) - \dfrac{ 2Mr_{-}}{r_{+}-r_{-}}\ln \left(  \dfrac{r-r_{-}}{2M}\right).
\end{displaymath}
We denote the position of the wall as $r_{*}^{0} \equiv r_*(r_0)$. The reflective properties of the wall are  parametrized by its reflectivity $R$, which accounts for the percentage of energy that is reflected back from the wall. 

\subsection{Green's function formulation}
\label{sec:green}

Our starting point is the time-independent radial Teukolsky equation \cite{Teukolsky1,Teukolsky2,Teukolsky3}, which is used for evolving field perturbations of spin $s$ around a Kerr black hole and is given by
\begin{align}
&\left( \dfrac{K^{2}-2\textit{i}s(r-M)K}{\Delta} + 4\textit{i}s\omega r - {}_{s}\lambda_{lmc} \right){}_{s}R_{lm\omega}(r) + \nonumber \\
&+ \Delta^{-s}\dfrac{d}{dr}\left( \Delta^{s+1}\dfrac{d {}_{s}R_{lm\omega}(r)}{dr} \right) = 0,
\label{eqteukolsky}
\end{align}
where $c\equiv a\omega$, $K \equiv (r^{2}+a^{2})\omega -am$ and ${}_{s}\lambda_{lmc}$ is the eigenvalue of the $(l,m)$ spin-weighted spheroidal harmonic ${}_{s}S_{lm\omega}$ \cite{Marcangular}. Therefore ${}_{s}R_{lm\omega}(r)$ is the radial part of a spin $s$ perturbation for a given $(l,m)$ in the frequency $\omega$ domain.

\begin{figure}
    \begin{center}
\resizebox{180pt}{90pt}{
    \begin{tikzpicture}
       \draw [ultra thick] (0,2.13) -- (2.13,0) ;
       \draw [ultra thick] (0,2.13) -- (2.13,4.26) ;
       \draw [ultra thick] (2.13,4.26) -- (4.26,2.13) ;
       \draw [ultra thick] (2.13,0) -- (4.26,2.13)  
node[below=80,left=28]{\Large{in-mode}};
      \draw [->,>=stealth, line width=2] (3.195,1.065) -- (1.065,3.195);
      \draw [->,>=stealth,  line width=2] (2.13,2.13) -- (3.195,3.195);
     \draw [ultra thick] (6,2.13) -- (8.13,0) ;
       \draw [ultra thick] (6,2.13) -- (8.13,4.26) ;
      \draw [ultra thick] (8.13,4.26) -- (10.26,2.13) ;
      \draw [ultra thick] (8.13,0) -- (10.26,2.13) 
node[below=80,left=28]{\Large{up-mode}};
      \draw [->,>=stealth,  line width=2] (8.13,2.13) -- (7.065,3.195);
      \draw [->, >=stealth, line width=2] (7.065,1.065) -- (9.195,3.195);
    \end{tikzpicture}}
  \end{center}
\caption{Penrose diagrams with schematic representation of the in and 
up-mode solutions of the radial Teukolsky equation, given by 
eqs.~(\ref{eqin}) and (\ref{equp}), respectively.}
\label{fig:penrose}
\end{figure}
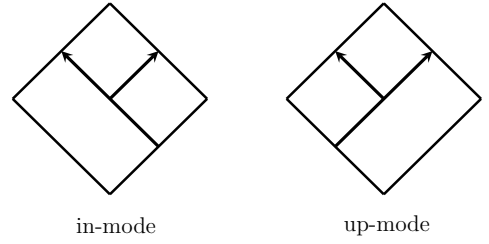

We now drop the index $s$ and consider only the scalar case $(s = 0)$. As usual, we need two independent solutions of the homogeneous equation (\ref{eqteukolsky}), which we choose as the \emph{in-} and \emph{up-going} solutions (see Figure \ref{fig:penrose}). Their asymptotic behavior reads: 
\begin{equation}\label{eqin}
R_{lm\omega}^{\rm in} \sim \left\lbrace \begin{array}{lr} 
B^{\rm trans}_{lm\omega}e^{-\textit{i}kr_{*}}, & \text{for} \quad r\rightarrow r_{+}\\ 
B^{\rm ref}_{lm\omega}\dfrac{e^{\textit{i}\omega r_{*}}}{r} + B^{\rm inc}_{lm\omega}\dfrac{e^{-\textit{i}\omega r_{*}}}{r}, &\text{for} \quad r\rightarrow \infty 
\end{array} \right.
\end{equation}
\begin{equation}\label{equp}
R_{lm\omega}^{\rm up} \sim \left\lbrace \begin{array}{lr} 
C^{\rm ref}_{lm\omega} e^{-\textit{i} k r_{*}} + C^{\rm inc}_{lm\omega}e^{\textit{i} k r_{*}}, & \text{for} \quad r\rightarrow r_{+} \\ 
C^{\rm trans}_{lm\omega}\dfrac{e^{\textit{i}\omega r_{*}}}{r}, & \text{for} \quad r\rightarrow \infty
\end{array} \right.
\end{equation}
 where $k \equiv \omega - m \Omega_{H}$ and $\Omega_{H}\equiv a/(2Mr_{+})$. In equations (\ref{eqin}) and (\ref{equp}) the coefficients $B^{\rm inc/ref/trans}_{lm\omega}$ and $C^{\rm inc/ref/trans}_{lm\omega}$ are the asymptotic amplitudes of the incident/reflected/transmitted parts of the ingoing and outgoing waves respectively and they are functions of the mode numbers $(l,m)$ and of the frequency $\omega$.

The Green's function of the radial Teukolsky equation (\ref{eqteukolsky}) can be written in terms of the homogeneous solutions (\ref{eqin}) and (\ref{equp}) as:
\begin{align}\label{GF}
G^{\rm BH}_{lm\omega}(r|r') =& \ \dfrac{R_{lm\omega}^{\rm up}(r)R_{lm\omega}^{\rm in}(r')}{W_{lm\omega}}\Theta(r-r')\nonumber +\\&+ \dfrac{R_{lm\omega}^{\rm up}(r')R_{lm\omega}^{\rm in}(r)}{W_{lm\omega}}\Theta(r'-r),
\end{align}
where $W_{lm\omega} = 2i \omega C^{\rm trans}_{lm\omega}B^{\rm inc}_{lm\omega} $ is the Wronskian of the two homogeneous solutions and $\Theta$ is the Heaviside function. 

In order to study the evolution of the perturbation and the development of echoes in an ECO spacetime, we follow \cite{Mark:2017dnq} and modify the Green's function to account for the new boundary condition in the ECO case. This modification results from the presence of the wall of reflectivity $R$ at a position $r_{*}^{0}$ and it is equivalent to replacing $R_{lm\omega}^{\rm in}(r)$ in (\ref{GF}) by $R_{lm\omega}^{\rm ECO}(r)$ given as 
\begin{equation}
R_{lm\omega}^{\rm ECO}(r) = R_{lm\omega}^{\rm in}(r)+ K_{lm\omega} R_{lm\omega}^{\rm up}(r), 
\end{equation}
where $K_{lm\omega} $ is the transfer function defined as 
\begin{eqnarray}
K_{lm\omega} &=& \frac{B^{\rm trans}_{lm\omega}}{C^{\rm trans}_{lm\omega}} \bar{K}_{lm\omega}\,, \quad {\rm where} \nonumber\\ 
\bar{K}_{lm\omega} &\equiv&  \frac{C^{\rm trans}_{lm\omega}}{C^{\rm inc}_{lm\omega}} \frac{R e^{-2i k r_{*}^{0}}}{1- (C^{\rm ref}_{lm\omega}/C^{\rm inc}_{lm\omega})Re^{-2i k r_{*}^{0}}},
\label{Komegabar}
\end{eqnarray}
and we denote $\bar{K}_{lm\omega}$ as the scaled transfer function. The expression for $K_{lm\omega}$ (\ref{Komegabar}) is computed by requiring that the asymptotic behavior near the wall
\begin{equation}
R_{lm\omega}^{\rm ECO}(r)\sim A^{\rm in}_{lm\omega}e^{-ikr_{*}}+A^{\rm out}_{lm\omega}e^{ikr_{*}}\,,
\end{equation}
has $A^{\rm out}_{lm\omega}=A^{\rm in}_{lm\omega}Re^{-2ikr^0_{*}}$, where $A^{\rm in/out}_{lm\omega}$ are the ingoing/outgoing amplitudes of the wave as $r\rightarrow r_{+}$.

Defined in this way, the new Green's function can be separated in two parts:
\begin{align}
G^{\rm ECO} &= G^{\rm BH} + G^{\rm echo}, \\
{\rm with} \quad G^{\rm echo} &= K_{lm\omega} \dfrac{ R_{lm\omega}^{\rm up}(r)R_{lm\omega}^{\rm up}(r')}{W_{lm\omega}},
\end{align}
where the first part reproduces the usual black hole response given by (\ref{GF}), that is, the black hole ringdown in QNMs. The second part generates the subsequent echoes. This implies that, if the reflectivity $R \to 0$, one should expect no deviations from the black hole waveform. 

We use as the initial excitation of the ECO waveform an infalling particle with scalar charge $q$ and energy density 
\begin{equation}
    \rho(x) = q\int d\tau \frac{\delta^{4}(x - x_{p}(\tau))}{\sqrt{-g}},
\end{equation}
where $x_{p}(\tau) = (t_p,r_p,\theta_p,\phi_p)(\tau)$ is the particle's trajectory as a function of its proper time $\tau$. We decompose $\rho(x)$ as 
\begin{equation}
\rho = \frac{1}{\Sigma}\sum_{lm} \int d\omega \rho_{lm\omega} {}_{s}S_{lm}(\theta) e^{im\phi}e^{-i\omega t},
\end{equation}
to select the appropriate $(l,m)$ mode and we find that the final echo response will be given by
\begin{equation}\label{phiecho}
\Phi_{lm\omega}(r)= \int G^{\rm echo}\rho_{lm\omega} dr' \equiv \bar{K}_{lm\omega}\dfrac{e^{i\omega r_{*}}}{r}Z^{\rm BH,H}_{lm\omega}, 
\end{equation}
where $Z^{\rm BH,H}_{lm\omega}$ is defined as the wave that would be entering the horizon in the black hole case, and is now being reflected back away from the horizon due to the presence of the wall. We can explicitly write this quantity as
\begin{align}\label{Z}
 & Z^{\rm BH,H}_{lm\omega}\propto q \frac{B^{\rm trans}_{lm\omega} }{2i\omega B^{\rm inc}_{lm\omega}C^{\rm trans}_{lm\omega}} \ \times  \nonumber\\
 & \times \int_{-\infty}^{\infty}d\tau R^{\rm up}_{lm\omega}(r_{p}){}_{0}S_{lmc}^{*}(\theta_{p})e^{-i m \phi_{p}}e^{i \omega t_{p}},
\end{align}
where we use the MST expression for $R^{\rm up}_{lm\omega}$ given in eq.~(\ref{417}) of Appendix \ref{MSTexplained}, switching to its asymptotic expansion for $r$ close to $r_{+}$.

\subsection{Orbits}
\label{sec:orbits}

It has long been known that an infalling particle excites the QNMs of a black hole \cite{PhysRevLett.27.1466} (see \cite{Berti_qnm_review} for a modern review). In \cite{Mark:2017dnq} it was argued that it is a reasonable choice to use a particle plunging from the innermost stable circular orbit (ISCO) as the source in the Teukolsky eq. (\ref{eqteukolsky}) to study the response of an ECO: in many cases binary systems will be quickly circularized  due to the emission of gravitational waves \cite{Peters1,Peters2,HadarIsco} and the ISCO is unique in the case of a non-spinning black hole. 

However, this is no longer the case when the central object is spinning. Assuming in-plane motion, in the spinning case there are two different ISCOs, one corating and one conterrotating with respect to the central object. In the nonrotating limit $a\rightarrow 0$ both of them will be located at $r=6M$, while in the extremal case $a\rightarrow M$  the counterrotating ISCO goes to $r=9M$, whereas the corotating ISCO approaches $r=r_{+}\to M$.

\begin{figure}
	   \includegraphics[width=0.45\textwidth]{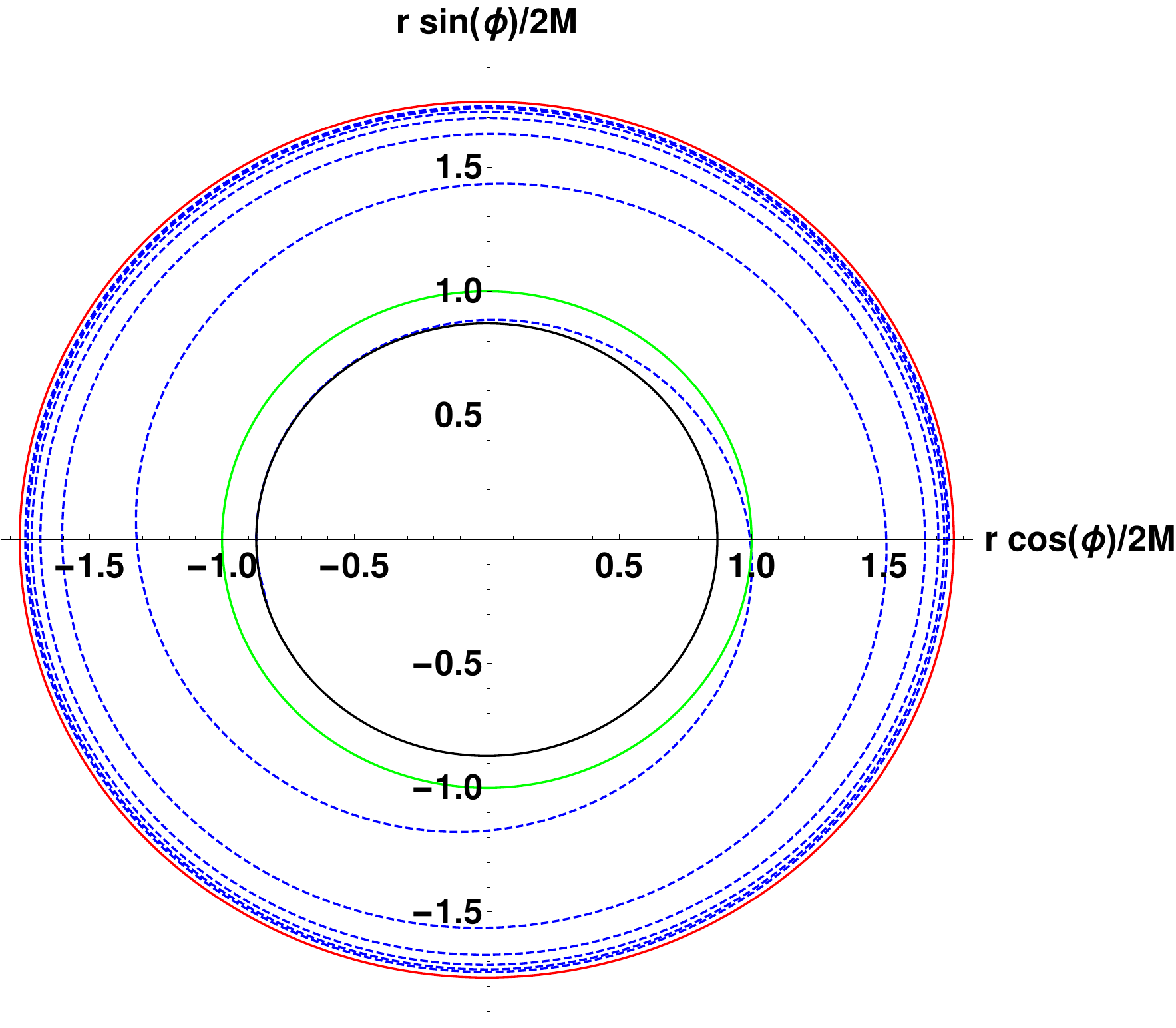}
	   \includegraphics[width=0.45\textwidth]{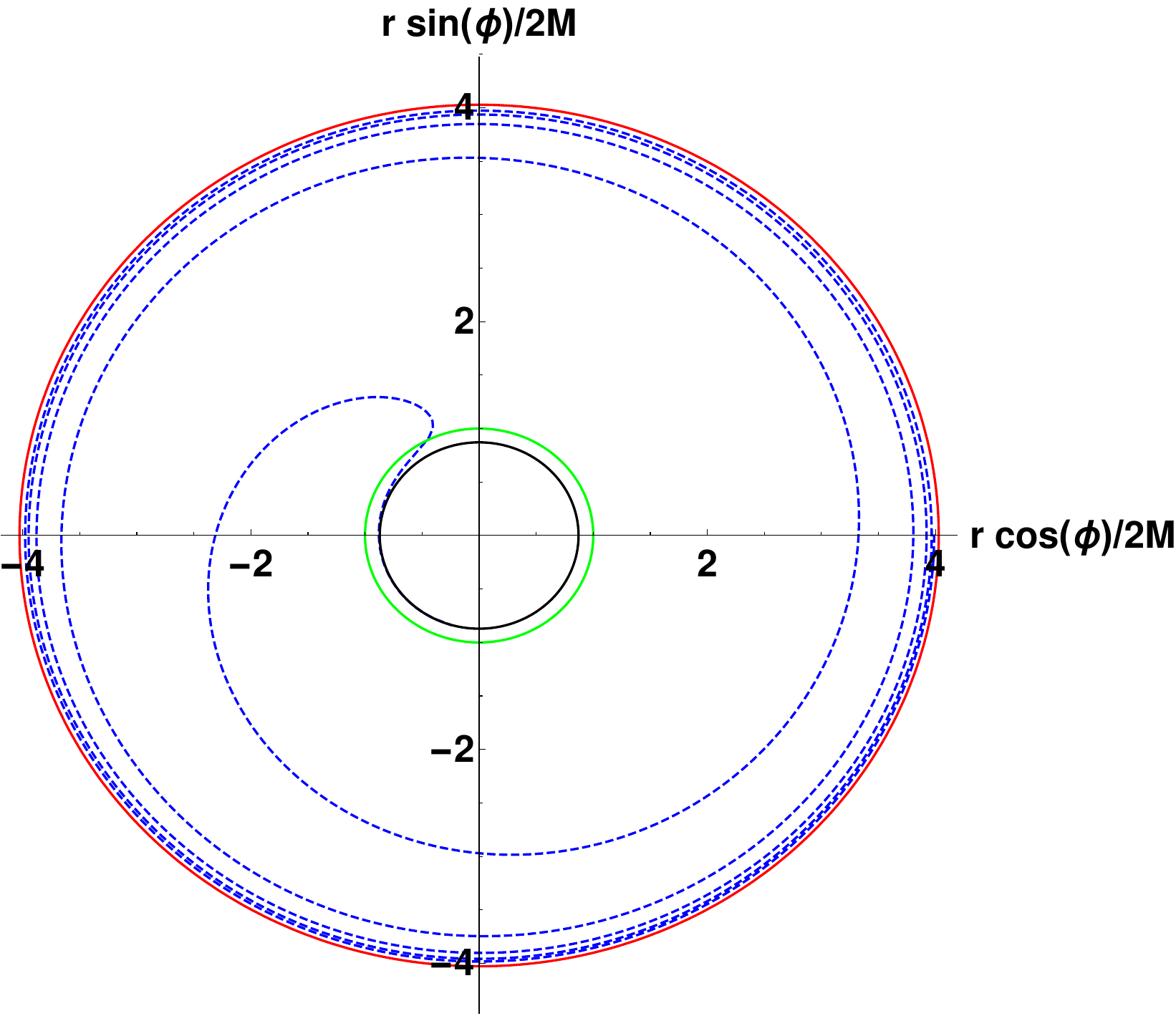}
\caption{Top: Corotating orbital trajectory of an infalling particle used for the evolution of the echo waveforms. The``would-be" horizon, the ergoregion and the ISCO radius are shown with black, green and red solid lines, respectively. The ISCO-plunge is shown with blue dashed lines, obtained from the integration of the geodesic equations of motion on the equatorial plane of a central object with spin parameter $a=0.67M$ and with initial position $r=0.99\,r_{\rm ISCO}$,  orbital energy $E=E_{\rm ISCO}$ and angular momentum $L=L_{\rm ISCO}$. Bottom: Same as the upper plot, but for a counterrotating orbital trajectory, which starts further away from the central object, spins down due to frame dragging and reverts its movement before it reaches the ergoregion.}
\label{orbits}
\end{figure}

In Figure \ref{orbits} we show two realizations of the geodesic motion that our test particle follows during the geodesic ISCO-plunge. Both orbits are in the equatorial plane ($\theta= \pi/2$) and the spin parameter of the central body is $a=0.67M$. The corotating orbit starts its motion closer to the central body than the counterrotating case. This could suggest that the perturbations from the corotating case would be stronger. However, as we discuss below in Section \ref{sec:K}, the transfer function is asymmetric in the rotating case, making it easier to excite counterrotating modes, and therefore the echoes excited in the counterrotating motion will have typically larger amplitudes (see Section \ref{echosandbeating}).

\subsection{Transfer function \label{Kbarr}}
\label{sec:K}
It is convenient to rewrite the scaled transfer function $\bar{K}_{lm\omega}$~(\ref{Komegabar}) using the transmission and reflection coefficients defined respectively as $T^{\rm BH}_{lm\omega}\equiv C^{\rm inc}_{lm\omega}/C^{\rm inc}_{lm\omega}$ and $R^{\rm BH}_{lm\omega}\equiv C^{\rm trans}_{lm\omega}/C^{\rm inc}_{lm\omega}$, which measure how much energy is transmitted to infinity and how much is reflected back to the black hole horizon in the up-going solution. Moreover, conservation of energy implies \cite{Teukolsky3}
\begin{equation}\label{conservationlaw}
\dfrac{\omega}{k} \dfrac{1}{2M r_{+}}|T^{\rm BH}_{lm\omega}|^{2} + |R^{\rm BH}_{lm\omega}|^2 = 1.
\end{equation}
In terms of the transmission and reflection coefficients, $\bar{K}_{lm\omega}$ takes the more simplified form 
\begin{equation}
\bar{K}_{lm\omega} =   \frac{T^{\rm BH}_{lm\omega}R e^{-2i k r_{*}^{0}}}{1- R^{\rm BH}_{lm\omega}Re^{-2i k r_{*}^{0}}}.
\end{equation}

\begin{figure}
	   \includegraphics[width=0.45\textwidth]{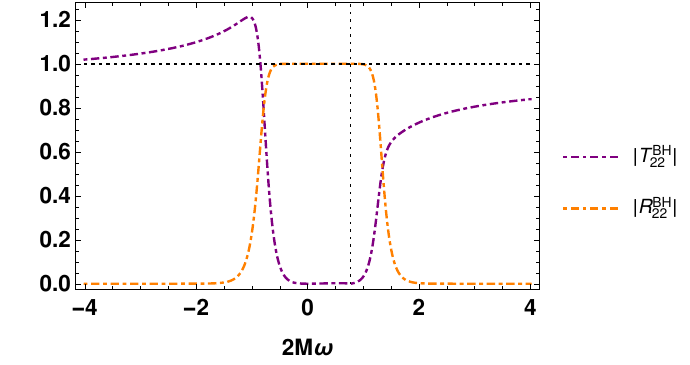}
	   \includegraphics[width=0.45\textwidth]{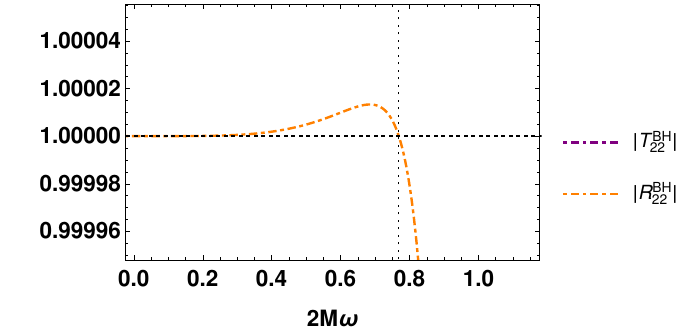}
	  \caption{Top: Absolute values of the transmission and reflection coefficients $|T^{\rm BH}_{lm}|$ and $|R^{\rm BH}_{lm}|$ as functions of the frequency $\omega$ for $l = m = 2$. The horizontal black line represents the energy conservation relation (\ref{conservationlaw}). The vertical dashed line marks the superradiant frequency $2M\omega_{\rm SR}\sim0.769$ for a Kerr black hole with $a = 0.67$. Bottom: Same as the upper plot, but zooming in on the superradiant frequencies. We will limit our choice of reflectivity $R$ to avoid the ergoregion instability in the ECO case.}
	  \label{conservationplot}
	\end{figure}

In Figure \ref{conservationplot} we show a numerical check of our method. The asymptotic solutions obtained with the MST expressions given in the Appendix \ref{MSTexplained} satisfy the conservation relation (\ref{conservationlaw}) to very good precision. 

\begin{figure}
	   \includegraphics[width=0.43\textwidth]{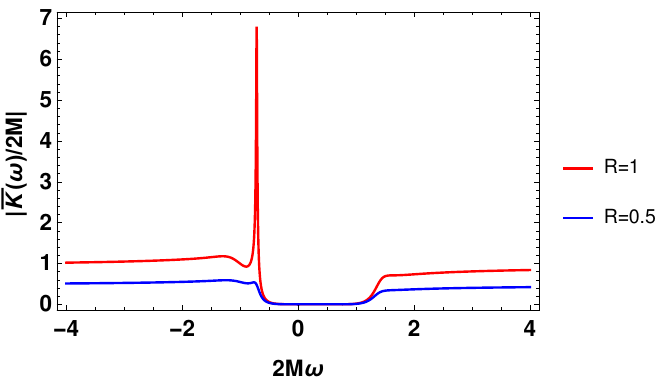}
	   \includegraphics[width=0.43\textwidth]{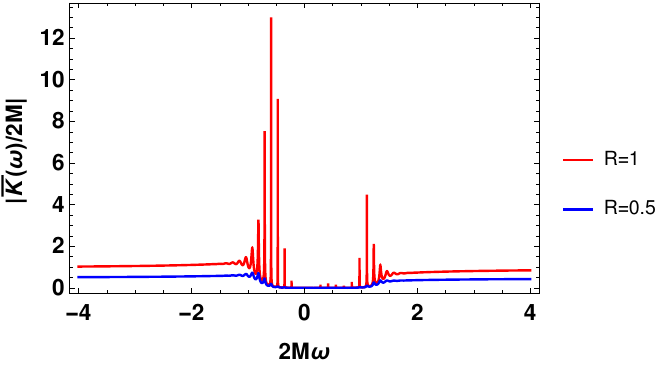}
\caption{Top: Absolute value of the scaled transfer function $\bar{K}_{lm\omega}$ as a function of the frequency $\omega$ for $l = m = 2$, for an ECO with $a = 0.67$, a reflective wall at $r_*^0 = -3M$ and two values of the reflectivity $R$. Bottom: Same as the upper plot, but for $r_*^0 = -50M$. The resonance peaks are more pronounced for more compact ECOs with stronger reflectivity, and their frequencies correspond to the QNMs of the ECO.}
\label{Kfull}
\end{figure}

The transfer function depends on the properties of the ECO. In Figure \ref{Kfull} we show our results for $\bar{K}_{lm\omega}$ for two different reflectivities ($R=0.5 , 1$) and two different wall positions ($r_{*}^{0}=-3M,-50M$). We note that lower reflectivities result in lower amplitudes of the observed waveform, as expected, and the resonances indicate the QNMs of the ECO.

The number of visible resonances in $\bar{K}_{lm\omega}$ increases as $r_{*}^{0}\rightarrow -\infty$, i.e, for more compact objects. Therefore we find it numerically convenient to work with the following expansions:
\begin{subequations}
    \begin{empheq}[left={\bar{K}_{lm\omega}=\empheqlbrace\,}]{align}
    \label{Kbarexpansion1}
      & \sum_{n=0}^{\infty} \dfrac{T^{\rm BH}_{lm\omega}}{(R^{\rm BH}_{lm\omega})^{-(n - 1)}}R^{n}e^{-2nikr_{*}^{0}},\nonumber\\
      & \quad\text{if}\ |R R^{\rm BH}_{lm\omega}|<1,\\
      \label{Kbarexpansion2}
      & \sum_{n=0}^{\infty} -\dfrac{T^{\rm BH}_{lm\omega}}{(R^{\rm BH}_{lm\omega})^{n}}R^{(-n + 1)}e^{2i(n - 1)kr_{*}^{0}},\nonumber\\
      & \quad\text{if}\ |R R^{\rm BH}_{lm\omega}|>1.   
    \end{empheq}
\end{subequations}
The expansion (\ref{Kbarexpansion1}) shows that each $n$-th term will generate the $n$-th echo in the signal, because the negative complex phases are responsible for shifting the signal to later times. We will not consider here cases when the expansion (\ref{Kbarexpansion2}) is applicable, that is, when the ergoregion instability is active and the resulting waveform is exponentially growing. Therefore, we  take $R$ such that $|R R^{\rm BH}_{lm\omega}|<1$ for all frequencies.\footnote{Though expansion (\ref{Kbarexpansion2}) is formally correct for the superradiant frequencies, it would be wrong to conclude that its positive complex phase indicates that the subsequent echoes appear \emph{earlier} in time. In this case, the ergoregion instability is active and the waveform is not square-integrable; hence the analysis in frequency space is not valid.}

\begin{figure}
	   \includegraphics[width=0.42\textwidth]{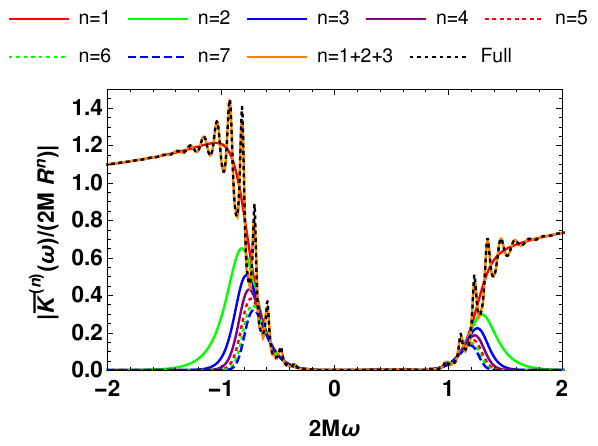}
	   \includegraphics[width=0.40\textwidth]{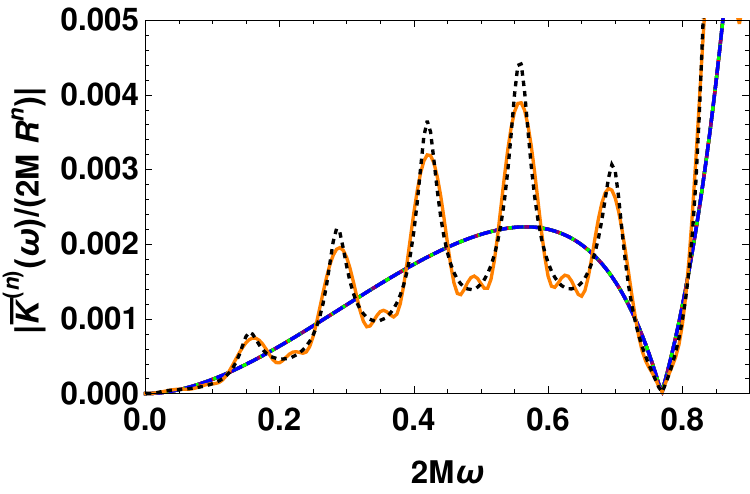}
\caption{Top: Absolute value of the first terms of expansion (\ref{Kbarexpansion1}) for the scaled transfer function $\bar K_{lm\omega}$ in the case of a rotating ECO with a = 0.67 and a wall with $R=0.5$ at $r_{*}^{0}=-50M$. The sum of the first three terms (in orange) is already very close to the full expression, shown in dotted black. Bottom: Same as the upper plot, but zooming in on the superradiant region ($0<2M\omega<2M\omega_{\rm SR}\sim0.769$). }
\label{Komegabarexpand}
\end{figure}

Figure \ref{Komegabarexpand} presents an example of expansion (\ref{Kbarexpansion1}), which converges rapidly as shown. The peaks come from the constructive interference of the complex phases for each term. Additionally, a small superradiant excess can be seen in all terms. This excess would be larger for the gravitational case and/or for a central object with higher spin (see for instance \cite{Casals:2019vdb}, and \cite{Brito:2015oca} for a review).

\section{Results for the waveforms}
\label{sect6}
\subsection{Horizon waveforms}
\label{sec:horizon}

The first step to obtain the waveform excited by a particle falling into the ECO (following the orbits described in Section \ref{sec:orbits}) is to construct the wave $Z^{\rm BH,H}_{lm\omega}$ that would reach the horizon, defined in eq.~(\ref{Z}).\footnote{Similarly to \cite{Mark:2017dnq}, we need to multiply the scalar charge in eq.~(\ref{Z}) by an ``activation function" to avoid the sudden appearance of the scalar particle near the ISCO.}

\begin{figure}
	   \includegraphics[width=0.38\textwidth]{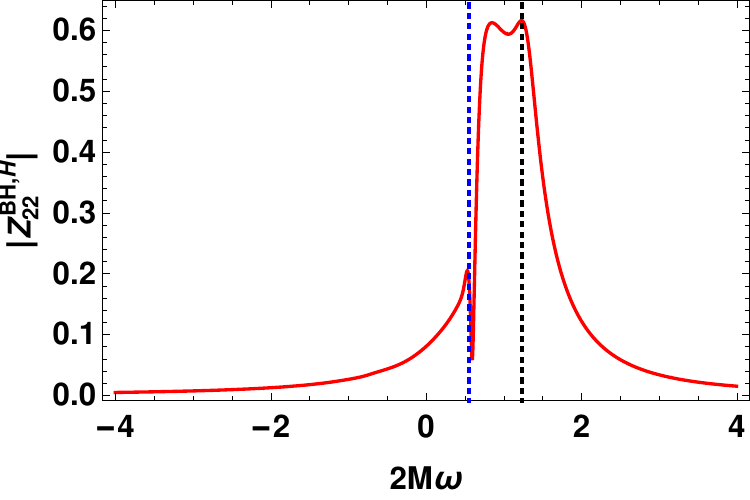}
	   \includegraphics[width=0.38\textwidth]{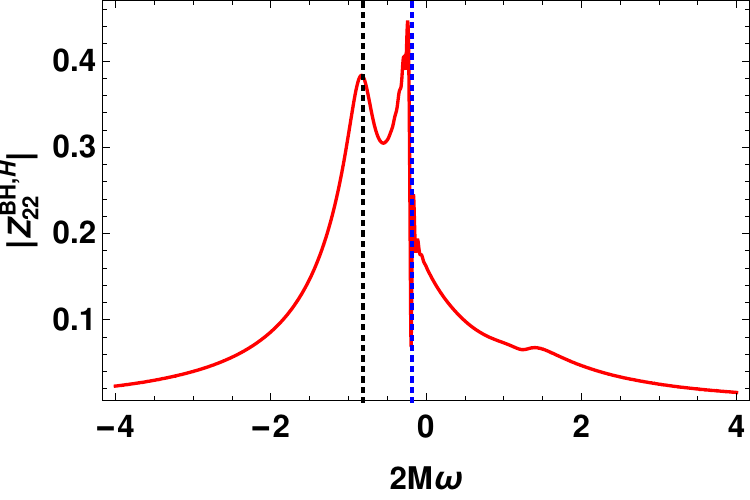}
\caption{Top: Absolute value of the horizon waveforms in the frequency domain for the corotating case. The black vertical dashed line marks the $s=0,l=m=2,n=2$ QNM of the black hole. Bottom: Same as the upper plot, but for the counterrotating case. The black vertical dashed line represents the $s=0,l=2,m=-2,n=0$ QNM of the black hole. In both plots the blue vertical dashed line represents the orbital frequency at the ISCO.}
\label{Zplotsomega}
\end{figure}

Figure \ref{Zplotsomega} shows our results for the corotating and counterrotating orbits presented in Figure \ref{orbits}.  In Figure \ref{Zplotsomega} the peaks marked in the upper and lower plots with dashed lines correspond exactly to the corotating $(l,m) = (2,2)$ and counterrotating $(l,m) = (2,-2)$ black hole QNM frequencies, respectively, which dominate the time-dependent waveform after the particle reaches the event horizon. At earlier times the waveforms are dominated by the contribution from the inspiral, with increasing frequency $\omega = 2\omega_{\rm orbital}$ starting close to twice the orbital frequency of the ISCO.

\subsection{Echo waveforms}
\label{echosandbeating}

Now that we have the scaled transfer function $\bar{K}_{lm\omega}$ (Section \ref{sec:K}) and the horizon waveform $Z^{\rm BH,H}_{lm\omega}$ (Section \ref{sec:horizon}), we can construct the echoes as prescribed by equation (\ref{phiecho}). We calculate each of the first echoes separately, using expansion (\ref{Kbarexpansion1}), summing them together in order to obtain the full echo waveform \cite{Mark:2017dnq}.

\begin{figure}
	   \includegraphics[width=0.38\textwidth]{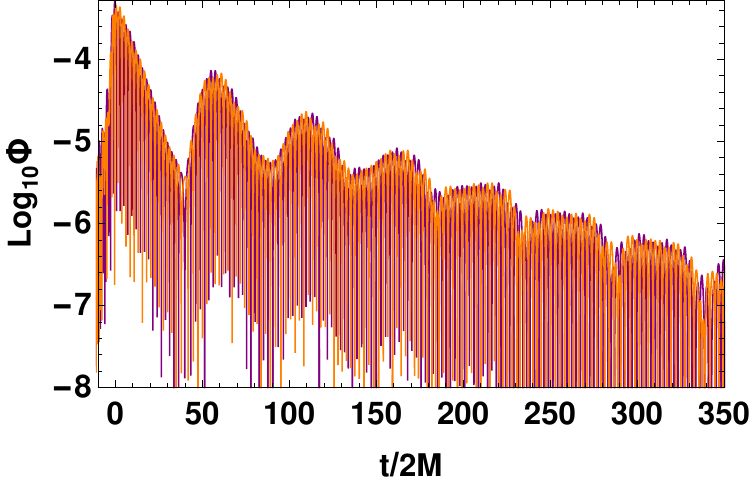}
	   \includegraphics[width=0.38\textwidth]{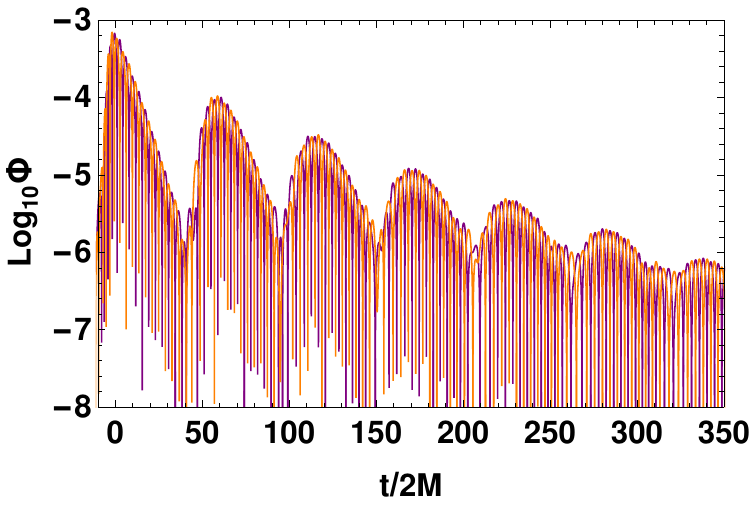}
\caption{Top: Real (orange) and imaginary (purple) parts of the echo waveform, excited by an infalling particle that was initially corotating with the ECO. The ECO is described by a reflective wall at $r^*_0 = 50$ with reflectivity $R = 0.5$, and the waveform was extracted at $150 M$. Bottom: Same as the upper plot, but for the counterrotating case. The vertical axis has units of $q/M$ and the time axis was shifted to set the peak of the echo waveform at $t=0$.}
\label{Echolog}
\end{figure}

We present our results for an ECO with representative values of spin $a = 0.67M$, reflectivity $R=0.5$ and position of the reflective wall $r_{*}^{0}=-50M$. For this choice of parameters $a$ and $R$, the ergoregion instability is suppressed (as it satisfies the condition $|RR^{BH}_{lm\omega}| < 1$, see also \cite{Paniquench}), and this position of the wall results in non-overlapping consecutive echoes. Our results are shown in Figure \ref{Echolog} for two cases, in which the infalling particle followed a trajectory from the corotating ISCO plunge (upper plot) or from the counterrotating ISCO plunge (lower plot). It is important to note that higher (constant) values of the reflectivity imply slower decay rates for subsequent echoes, without altering their frequencies, see eq.~(\ref{Kbarexpansion1}).

\begin{figure*}
	   \includegraphics[width=0.32\textwidth]{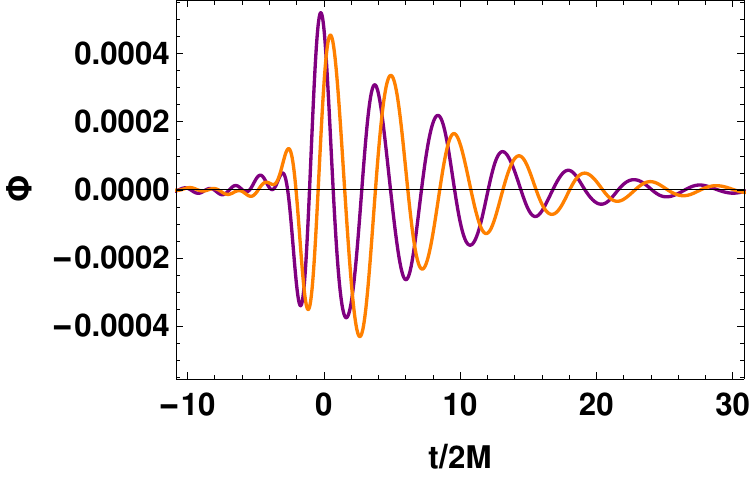}
	   \includegraphics[width=0.32\textwidth]{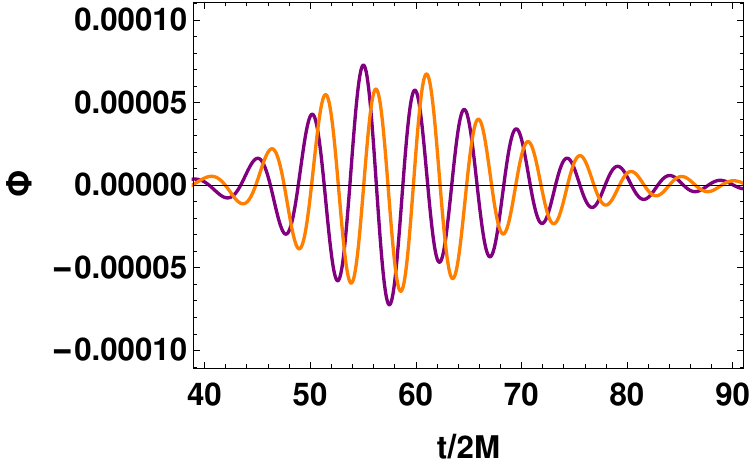}
	   \includegraphics[width=0.32\textwidth]{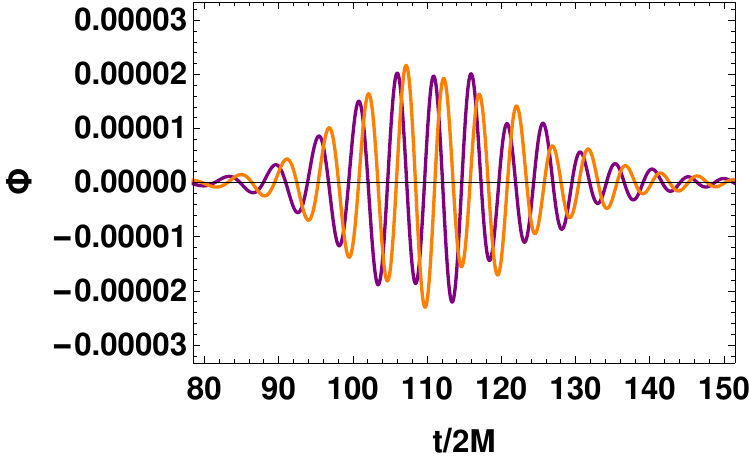}
	   \includegraphics[width=0.32\textwidth]{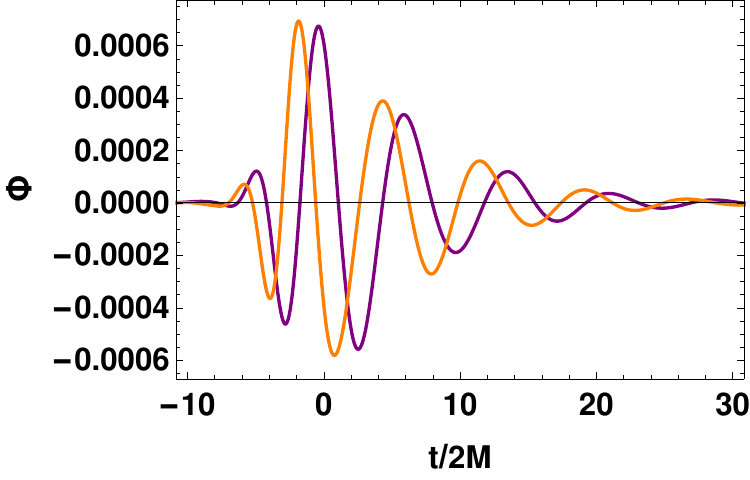}
	   \includegraphics[width=0.32\textwidth]{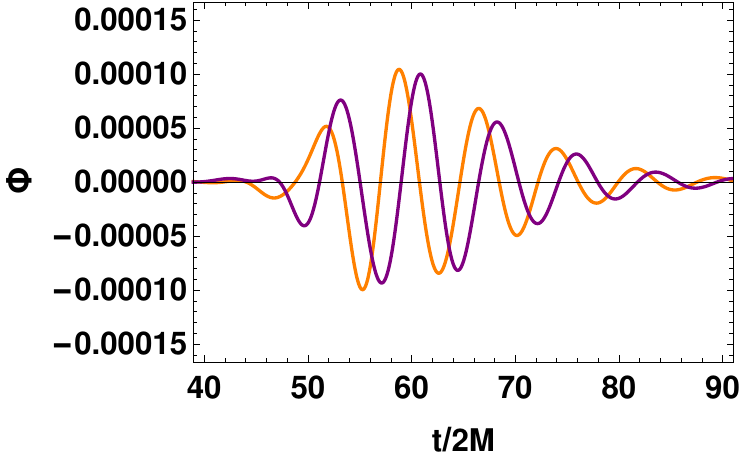}
	   \includegraphics[width=0.32\textwidth]{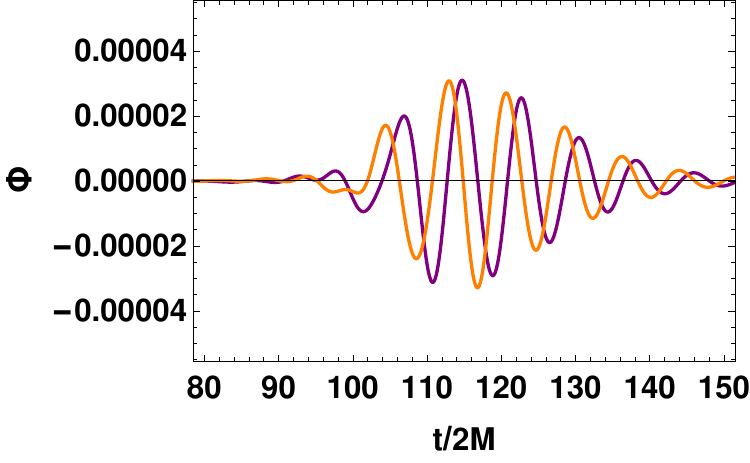}
\caption{Same as Figure \ref{Echolog}, but showing the first echoes separately in linear scale. From left to right we have the first, second and third echoes. On the top (bottom) row, the echoes for the corotating (counterotating) case. The echoes in the corotating case have amplitudes approximately 30\% smaller and frequencies approximately 40\% larger than the counterrotating case.}
\label{Echon1}
\end{figure*}

In Figure \ref{Echon1} we show separately from left to right the first, second and third echoes of the waveforms presented in Figure \ref{Echolog}. We can see that the period of the oscillations is shorter in the echoes resulting from the plunge of the corotating particle (shown in the upper plots) and longer in the counterrotating case (lower plots). This result can be partially understood noting that the corotating particle is spun up by frame dragging, while the counterrotating particle spins down until it reverses its movement (see Figures \ref{orbits} and \ref{Zplotsomega}). The amplitudes of the corresponding echoes are similar, but the counterrotating case has higher amplitudes due to the asymmetry of the scaled transfer function $\bar{K}$ (see Figure \ref{Kfull}).\footnote{However, it would be possible to fine tune the transfer function to select the same frequencies for both orbits, as in the nonrotating case \cite{Mark:2017dnq}. A simple example of such a transfer function would be a sum of two gaussians centered at $\pm \omega_{0}$} 

\begin{figure*}
\includegraphics[width=0.32\textwidth]{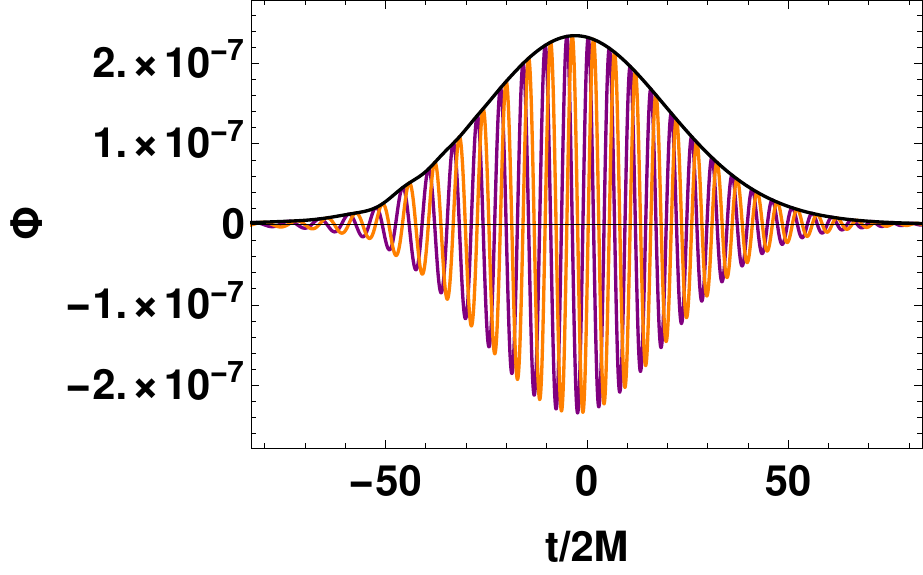}
\includegraphics[width=0.32\textwidth]{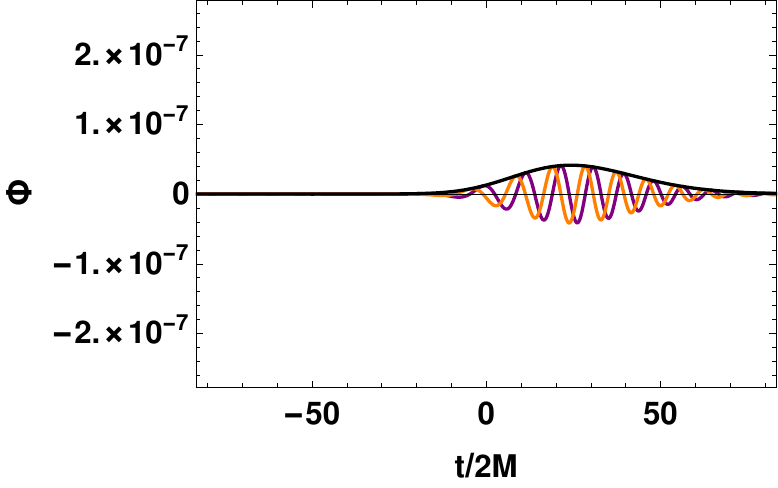}
\includegraphics[width=0.32\textwidth]{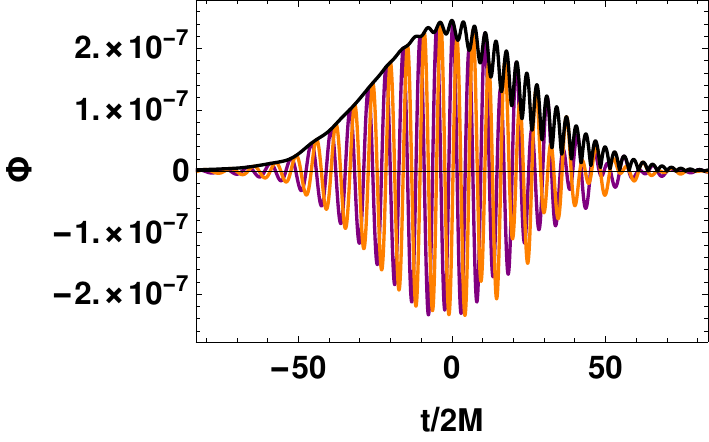}
\includegraphics[width=0.32\textwidth]{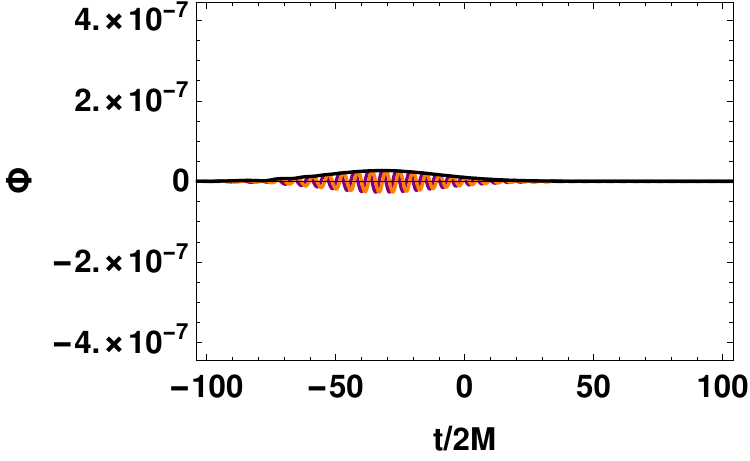}
\includegraphics[width=0.32\textwidth]{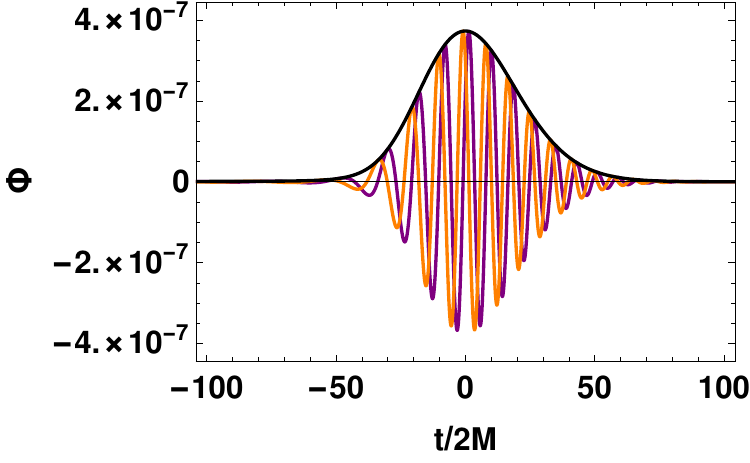}
\includegraphics[width=0.32\textwidth]{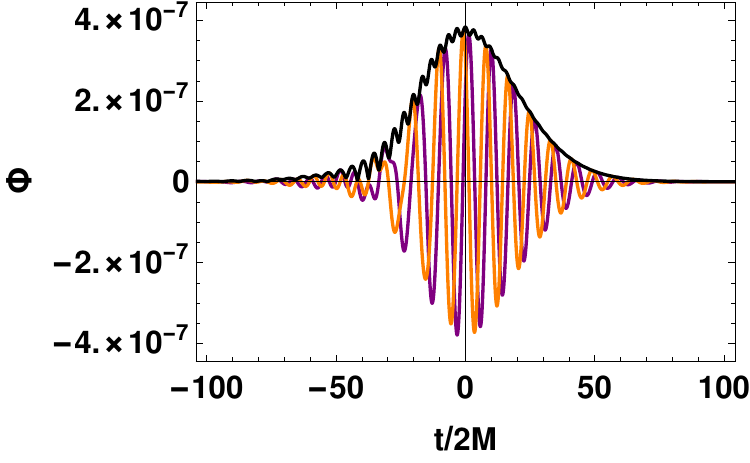}
\caption{Same as Figure \ref{Echon1}, but for 8th echo, showing the positive and negative frequency contributions separately in the left and center plots, respectively. The complete echo is shown on the right plots. In the corotating case (top row), the subdominant contribution comes from the negative frequency contribution. The situation is reversed for the counterrotating case (bottom row). The black curve indicates the amplitude of the complex waveform and shows how the superposition of the two approximately gaussian pulses on the left and center results in the beating of the full waveform on the right.}\label{decompose}
\end{figure*}

It is also interesting to note that each echo has a subdominant frequency contribution, which can be seen in the beating of the waveform that becomes more pronounced for later echoes.  We present in Figure \ref{decompose} the explicit example of the 8th echo for both the corotating (top row) and the counterrotating (bottom row) cases. We have isolated the positive and negative frequency contributions in the left and center plots, respectively, and the full echo showing the beating is in the right plots. This effect is also caused by the rotation of the ECO, which breaks the symmetry between positive and negative frequencies in the non-rotating case.\footnote{An unrelated modulation of the waveform was reported in  \cite{PaniAnalytical}, where it was noticed that rotation can mix the $+$ and $\times$ polarizations in the gravitational case.}

A preliminary analysis at a higher spin a = 0.9 indicated, as expected, that the asymmetry between positive and negative frequencies increases with the ECO spin. However, energy conservation (given by eq. (\ref{conservationlaw})) limits the amplitude of each frequency contribution. Therefore, a higher spin results in more easily resolvable negative and positive frequency contributions, but not in a stronger contribution from the secondary frequency.

Consequently, the resulting waveform can be approximately described by the superposition of two gaussian pulses with distinct frequencies. Therefore it is natural to propose an extension of the parametrization of the echo waveform as \emph{one} gaussian pulse proposed in \cite{Cardosomorphology}, in order to also account for the subdominant \emph{second} pulse described here. 

However, it is important to note that such a modification would necessarily need 14 independent parameters, nearly doubling the initial 8 parameters of the single guassian model (see eq.~(2) in \cite{Cardosomorphology}). This indicates the increasingly complex nature of the echoes in the rotating case, which should also make the ECO parameter estimation \cite{Kokkotasechoesparameter} harder to perform. 
It has also been alternatively suggested \cite{HoldomWaves} that a search for the characteristic resonance peaks (see for example Figure \ref{Kfull}) could be a more promising way to obtain information on the parameters of the ECO.

\section{Conclusions}
\label{sect7}

The detections of gravitational waves from compact binary coalescences allow us to test general relativity and the very existence of black holes, probing gravity in the strong field regime close to compact objects. If classical black holes are not formed in nature, for instance due to quantum corrections that could prevent the formation of an event horizon, the gravitational wave signal from these events could present non-trivial modifications.

We have extended the method first developed in \cite{Mark:2017dnq} to study scalar perturbations around a \emph{rotating} ECO. The extension to the rotating case is crucial for obtaining results that can be compared with observational data. We used the MST method for obtaining the homogeneous solutions of the Teukolsky equation and a Green's function formulation to calculate the echoes with an infalling particle as the source. 
Our source was allowed to follow two different paths: a corotating and a counterrotating ISCO plunge. The resulting echoes show different characteristics; the corotating case has an amplitude about 30\% smaller and the the frequencies are approximately 40\% larger than the counterrotating case. 

The rotation of the ECO also introduces a subdominant frequency contribution, which is responsible for the beating of the waveform observed more easily in later echoes. Here we were able to find a simple way to describe this modulation as a sum of two independent gaussian wave packets. 

We are currently working on the generalization of our numerical setup to the gravitational case, in which we expect that the main features observed in the scalar case will be preserved. Therefore we propose a simple modification of the single gaussian template given in \cite{Cardosomorphology}, which we expect  will enhance the chances of detection, perhaps by stacking the data from different events \cite{HuanYangStacking,2018PhRvL.120c1102B}.

\begin{acknowledgments}
We thank Niayesh Afshordi, Vitor Cardoso, Bob Holdom, Cole Miller, Paolo Pani, Luciano Rezzolla, Maur\'icio Richartz and Alan Maciel da Silva, for useful discussions and comments. LFLM was supported in part by grant 2017/24919-4  of the S\~ao Paulo Research Foundation (FAPESP) and by the Coordena\c c\~ao de Aperfei\c coamento de Pessoal de N\'ivel Superior - Brasil (Capes) - Finance code 001 through the Capes-PrInt program. CC acknowledges support from grant 303750/2017-0 of the Brazilian National Council for Scientific and Technological Development (CNPq), from the Simons Foundation through the Simons Foundation Emmy Noether Fellows Program at Perimeter Institute and by NASA under award number 80GSFC17M0002. We are grateful for the hospitality of Perimeter Institute where part of this work was carried out. 
%
Research at Perimeter Institute is supported in part by the government
of Canada through the Department of Innovation, Science and Economic
Development and by the Province of Ontario through the Ministry
of Colleges and Universities.

\end{acknowledgments}

\appendix

\section{MST Method}
\label{MSTexplained}

We solve the radial Teukolsky equation with the MST method, developed in 1996 by Mano, Suzuki and Takasugi \cite{MSToriginal, SasakiReview}, which allows us to analytically calculate solutions for the homogeneous radial Teukolsky equation and their asymptotic amplitudes. Here we present the main results and some useful formulas from the MST method used in our work.

Solutions of the Teukolsky equation must satisfy the relation $R_{lm\omega}= \bar{R}_{l,-m,-\omega} $, where the bar denotes complex conjugation. Due to this property we can assume $\omega > 0$ to obtain the following equations and use this symmetry relation for $\omega < 0$.

With the MST method we obtain the two linearly independent solutions of the Teukolsky equation 
as an infinite sum of hypergeometric functions and irregular confluent hypergeometric functions. The solutions in which we are interested are given by:
\begin{widetext}
\begin{align} \label{416}
& R^{\rm in}_{lm\omega} =e^{i \epsilon \kappa y}(-y)^{-s-i(\epsilon+\tau)/2}(1-y)^{i(\epsilon-\tau)/2}\sum_{n=-\infty}^{\infty} \textit{a}_{n}{}_{2}F_1(n+\nu+1-i\tau,-n-\nu-i\tau;1-s-i\epsilon-i\tau;y), 
\end{align}
%
%
\begin{multline} \label{417}
	 R^{\rm ref}_{lm\omega} = 2^{\nu}e^{-\pi \epsilon}e^{-\textit{i}\pi(\nu+1+s)}
	e^{\textit{i}\hat{z}}\hat{z}^{\nu+\textit{i}\epsilon_+}(\hat{z}-\epsilon\kappa)^{-s-\textit{i}\epsilon_+} \\ \times\sum_{n=-\infty}^{\infty}i^{n}
	\frac{(\nu+1+s-\textit{i}\epsilon)_n}{(\nu+1-s+\textit{i}\epsilon)_n}
	\textit{a}_{n}(2\hat{z})^{n} U(n+\nu+1+s-\textit{i}\epsilon,2n+2\nu+2;-2\textit{i}\hat{z}),
\end{multline}
\end{widetext}
where $\epsilon \equiv 2M\omega$, $q\equiv a/M$, $\kappa \equiv \sqrt[]{1-q^{2}}$, $\tau \equiv (\epsilon - mq)/\kappa$, $y \equiv \omega(r_{+}-r)/(\epsilon \kappa)$, $\hat{z} \equiv \epsilon \kappa (1-y)$,  $\epsilon_+ \equiv (\epsilon+\tau)/2$, $(y)_{n}$ is the Pochhammer symbol and $\nu$ is a function of $(l,m,\omega)$ \cite{SasakiReview}. 

Both series (\ref{416}) and (\ref{417}) have the same coefficients $a_n$, which satisfy the three term recurrence relation:
\begin{equation} \label{418}
\alpha_n^{\nu} \textit{a}_{n+1}+\beta_n^{\nu} \textit{a}_{n}+\gamma_n^\nu \textit{a}_{n-1}=0, \quad \text{for} \quad  n \in \mathbb{Z},
\end{equation} 
with the following recurrence coefficients 
\begin{align} \label{mstcoeff}
\alpha_{n}^{\nu} &=\frac{i\epsilon\kappa(n+\nu+1+s+i\epsilon)(n+\nu+1+s-i\epsilon)}{(n+\nu+1)(2 n+2 \nu+3)(n+\nu+1+i\tau)^{-1}},  \\
\beta_{n}^{\nu} &=-{}_{s}\lambda_{lmc}-s(s+1)+(n+\nu)(n+\nu+1)\nonumber\\ &\quad\quad+\epsilon^2+\epsilon(\epsilon-m q)+\frac{\epsilon(\epsilon-m q)(s^2+\epsilon^2)}{(n+\nu)(n+\nu+1)}, \\
\gamma_{n}^{\nu} &=-\frac{i\epsilon\kappa(n+\nu-s+i\epsilon)(n+\nu-s-i\epsilon)(n+\nu-i\tau)}{(n+\nu)(2 n+2 \nu-1)}.
\end{align} 
The parameter $\nu$ is such that $\textit{a}_{n}$ is the minimal solution of the three-term recurrence relation both for $n\rightarrow -\infty$ and $n\rightarrow \infty$ (for a more comprehensive discussion we direct the reader to a review of the method in \cite{SasakiReview}). In our work $\nu$ was calculated following the prescription described in \cite{Castromonodromy} and available on \cite{Castromonodromy2}. The computation of the angular eigenvalues ${}_{s}\lambda_{lmc}$ and eigenfunctions ${}_{s}S_{lm\omega}$ used the toolkit available in \cite{toolkit}. 

In this construction of the homogeneous solutions we implicitly choose a particular normalization. In our choice the  in-mode coefficients are given by:
\begin{equation} \label{429}
B^{\rm trans}_{lm\omega}= \left(\dfrac{\epsilon \kappa}{\omega}\right)^{2 s}e^{i \kappa \epsilon_+(1+2\log\kappa/(1+\kappa))}\sum_{n=-\infty}^{\infty} \textit{a}_{n},
\end{equation}
\begin{eqnarray} \label{binc}
B^{\rm inc}_{lm\omega}
= &\omega^{-1}\left({K}_{\nu}-
\textit{i}e^{-\textit{i}\pi\nu} \frac{\sin \pi(\nu-s+\textit{i}\epsilon)}
{\sin \pi(\nu+s-\textit{i}\epsilon)}
K_{-\nu-1}\right)\times \nonumber \\ & \times A_{+}^{\nu} e^{-\textit{i}(\epsilon\ln\epsilon -\frac{1-\kappa}{2}\epsilon)},
\end{eqnarray}
\begin{equation}  \label{bref}
B^{\rm ref}_{lm\omega} = \omega^{-1-2s}\left(K_{\nu}+\textit{i}e^{\textit{i}\pi\nu} K_{-\nu-1}\right)A_{-}^{\nu} e^{\textit{i}(\epsilon\ln\epsilon -\frac{1-\kappa}{2}\epsilon)}.
\end{equation}

One of the up-going coefficients is given by: 
\begin{equation}
C^{\rm trans}_{lm\omega} = \omega^{-1-2s}A_{-}^{\nu} e^{\textit{i} (\epsilon \log\epsilon -\frac{1-\kappa}{2}\epsilon )},
\label{ctrans}
\end{equation}
and the remaining coefficients $C^{\rm ref}_{lm\omega}$ and $C^{\rm inc}_{lm\omega}$ can be found with Wronskian relations \cite{Marcextreme}. In order to find $C^{\rm inc}_{lm\omega}$ we use the conservation of the Wronskian $W_1$ defined as:
\begin{equation}
W_{1}=\Delta^{s+1} \left( R^{\rm up}_{lm\omega} \dfrac{d R^{\rm up}_{lm\omega}}{dr}- R^{\rm up}_{lm\omega}\dfrac{d  R^{\rm in}_{lm\omega}}{dr}\right),
\end{equation}
from which we find the relation:
\begin{equation}\label{Cinc}
C^{\rm inc}_{lm\omega} =\dfrac{\textit{i}\omega}{M}\left(2 \textit{i}k r_{+} +s \kappa \right)^{-1}\dfrac{B^{\rm inc}_{lm\omega}}{B^{\rm trans}_{lm\omega}} C^{\rm trans}_{lm\omega}.
\end{equation}
Similarly for finding $C^{\rm ref}_{lm\omega}$ we use the conservation of the Wronskian $W_2$ given by:
\begin{align}\label{W2}
W_{2} & =\Delta^{s+1} \left( R^{\rm in}_{s,lm\omega} \dfrac{d(\Delta^{-s}\bar{R}^{\rm up}_{-s,lm\omega})}{dr} 
-\right.\nonumber \\
&\left.
- \Delta^{-s} \bar{R}^{\rm up}_{-s,lm\omega}\dfrac{d  R^{\rm in}_{s,lm\omega}}{dr}\right),
\end{align}
from where we can find the last coefficient to be given as:
\begin{equation}\label{Cref}
C^{\rm ref}_{s,lm\omega} =\dfrac{\textit{i}\omega}{M}\left(-2 \textit{i}k r_{+}-s \kappa \right)^{-1}\dfrac{\bar{B}^{\rm ref}_{-s,lm\omega}}{\bar{B}^{\rm trans}_{-s,lm\omega}} C^{\rm trans}_{s,lm\omega}.
\end{equation}
Finally, the quantities $A_{+}^{\nu}$, $A_{-}^{\nu}$ and $K_{\nu}$ in eqs.~(\ref{binc}), (\ref{bref}) and (\ref{ctrans}) are defined as \cite{SasakiReview}:
\begin{widetext}
\begin{align} \label{427}
 A_{+}^{\nu}&\equiv e^{-{\pi\over 2}\epsilon}e^{{\pi\over 2}i(\nu+1-s)}
2^{-1+s-\textit{i}\epsilon}{\Gamma(\nu+1-s+\textit{i}\epsilon)\over 
\Gamma(\nu+1+s-\textit{i}\epsilon)}\sum_{n=-\infty}^{+\infty}\textit{a}_{n},\\
A_{-}^{\nu} &\equiv 2^{-1-s+\textit{i}\epsilon}e^{-{\pi\over 2}\textit{i}(\nu+1+s)}e^{-{\pi\over 2}\epsilon} \times\sum_{n=-\infty}^{+\infty}(-1)^n{(\nu+1+s-i\epsilon)_n\over 
(\nu+1-s+\textit{i}\epsilon)_n}\textit{a}_{n} ,
\end{align}
and
\begin{align} \label{428}
K_{\nu} \equiv &	\frac{e^{i\epsilon\kappa}(2\epsilon \kappa )^{s-\nu-\eta}2^{-s}i^{\eta}
	\Gamma(1-s-2\textit{i}\epsilon_+)\Gamma(\eta+2\nu+2)}
	{\Gamma(\eta+\nu+1-s+\textit{i}\epsilon)
	\Gamma(\eta+\nu+1+\textit{i}\tau)\Gamma(\eta+\nu+1+s+\textit{i}\epsilon)}\times \left(\sum_{n=-\infty}^{\eta}
	\frac{(-1)^n}{(\eta-n)!
	(\eta+2\nu+2)_n}\frac{(\nu+1+s-\textit{i}\epsilon)_n}{(\nu+1-s+\textit{i}\epsilon)_n}
	\textit{a}_{n} \right)^{-1}
	\nonumber\\
	&\times \left ( \sum_{n=\eta}^{\infty}
	(-1)^n\, \frac{\Gamma(n+\eta+2\nu+1)}{(n-\eta)!}
	\frac{\Gamma(n+\nu+1+s+\textit{i}\epsilon)}{\Gamma(n+\nu+1-s-\textit{i}\epsilon)}
	\frac{\Gamma(n+\nu+1+\textit{i}\tau)}{\Gamma(n+\nu+1-\textit{i}\tau)}
	\,\textit{a}_{n} \right),
	\nonumber\\
	&
\end{align}
\end{widetext}
where $\eta$ is an arbitrary integer which we set to zero in our code without loss of generality. These expressions allow us to calculate the asymptotic amplitudes directly.

\bibliography{scalar_echoes_paper}

\end{document}